\documentclass[times,twocolumn,final]{elsarticle}

\usepackage{medima}
\usepackage{framed,multirow}

\usepackage{amssymb}
\usepackage{latexsym}

\usepackage{url}

\usepackage{hyperref}

\usepackage{amsmath,amsfonts}
\usepackage{algorithmic}
\usepackage{graphicx}
\usepackage{textcomp}
\usepackage{caption}
\usepackage{subcaption}
\usepackage{verbatim} 
\usepackage{multirow}
\usepackage{xcolor}

\definecolor{newcolor}{rgb}{.8,.349,.1}

\journal{Computer Methods and Programs in Biomedicine}

\begin{document}

\verso{Yanbin Liu \textit{et~al.}}

\begin{frontmatter}

\title{Inflating 2D Convolution Weights for Efficient Generation of 3D Medical Images} 

\author[1]{Yanbin Liu\fnref{fn1}}
\fntext[fn1]{Most work was done at Harry Perkins Institute of Medical Research.}
\author[2,3]{Girish Dwivedi}
\author[4]{Farid Boussaid}
\author[5]{Frank Sanfilippo}
\author[6]{Makoto Yamada}
\author[7]{Mohammed Bennamoun\corref{cor1}}
\cortext[cor1]{Corresponding author. \\
 \indent \quad \emph{E-mail address}: mohammed.bennamoun@uwa.edu.au (M. Bennamoun).}

\address[1]{School of Computing, Australian National University{, Canberra, ACT, AU}}
\address[2]{Harry Perkins Institute of Medical Research, The University of Western Australia, Perth, WA, AU}
\address[3]{{Cardiology Department, }Fiona Stanley Hospital, Perth, WA, AU}
\address[4]{Department of Electrical, Electronic and Computer Engineering, The University of Western Australia, Perth, WA, AU}
\address[5]{School of Population and Global Health, The University of Western Australia, Perth, WA, AU}
\address[6]{Okinawa Institute of Science and Technology, Okinawa, JP}
\address[7]{Department of Computer Science and Software Engineering, The University of Western Australia, Perth, WA, AU}


\begin{abstract}
\emph{Background and Objective}:
The generation of three-dimensional (3D) medical images has great application potential since it takes into account the 3D anatomical structure. 
Two problems prevent effective training of a 3D medical generative model: (1) 3D medical images are expensive to acquire and annotate, resulting in an insufficient number of training images, and (2) a large number of parameters are involved in 3D convolution. \\
\emph{Methods}: 
We propose a novel GAN model called \emph{3D Split$\&$Shuffle-GAN}.
To address the 3D data scarcity issue, we first pre-train a two-dimensional (2D) GAN model using abundant image slices and inflate the 2D convolution weights to improve the initialization of the 3D GAN. 
Novel 3D network architectures are proposed for both the generator and discriminator of the GAN model to significantly reduce the number of parameters while maintaining the quality of image generation. 
Several weight inflation strategies and parameter-efficient 3D architectures are investigated. \\
\emph{Results}: Experiments on both heart (Stanford AIMI Coronary Calcium) and brain (Alzheimer’s Disease Neuroimaging Initiative) datasets show that our method leads to improved 3D image generation quality (14.7 improvements on Fréchet inception distance) with significantly fewer parameters (only 48.5\% of the baseline method). \\
\emph{Conclusions}: We built a parameter-efficient 3D medical image generation model. Due to the efficiency and effectiveness, it has the potential to generate high-quality 3D brain and heart images for real use cases.

\end{abstract}


\end{frontmatter}


\section{Introduction}
\label{sec:introduction}

With the availability of large-scale annotated datasets like ImageNet~\citep{ref:imagenet}, 
convolution neural networks (CNNs) have achieved unprecedented success in computer vision~\citep{ref:alexnet}. 
Benefiting from CNNs, medical imaging research has made great advancements in the classification~\citep{ref:medicalGANs4}, segmentation~\citep{ref:3DStylePET,ref:cmpb_cardiac,ref:cmpb_segment}, detection~\citep{ref:medicalDetect}, reconstruction~\citep{ref:cmpb_reconstruction}, and registration~\citep{ref:medicalRegistration} of two-dimensional (2D) medical images. 
However, 3D medical image research lags behind due to the lack of large-scale 3D medical image datasets. As a result of the complex collection procedure, expert annotation, privacy concerns and patient consent, it is challenging to build a large-scale, 3D medical dataset similar to ImageNet. 

One widely-used solution for the data deficit of medical images is Generative Adversarial Networks (GANs)~\citep{ref:GANs}.
These networks create high-quality synthetic images to mimic realistic data distributions. 
An example is using GANs with Wasserstein distance and perceptual loss for {low-dose computed tomography (CT)} image denoising~\citep{ref:medgans1}. 
Perceptual loss cannot be directly used for 3D medical images due to the lack of interpretable pre-trained 3D models. 
A cyclic loss GAN was used by~\cite{ref:medgans2} to reconstruct MRI images. 
Using cycle-consistent GANs, \cite{ref:medgans3} translated {magnetic resonance (MR)} images to CT images. 
Albeit effective in mitigating the data-deficit challenge, most existing GANs{-}based methods are designed for 2D medical image generation. 
Therefore, they do not incorporate information about the 3D anatomical structure~\citep{ref:3d_review}. 
Various medical applications require the 3D anatomical structure, including calcium scoring~\citep{ref:calcium1,ref:cmpb_calcium} of cardiac CT Coronary Angiograms (CTCAs), and brain tumor segmentation~\citep{ref:tumor,ref:cmpb_tumor}. 
Unfortunately, there are two practical issues that hinder the effective training of the 3D medical generative model, preventing the use of GANs in 3D medical imaging. 

First of all, there are usually insufficient 3D medical images to train effective 3D generative models. 
The effective training of 3D CNNs with natural videos relies on large-scale datasets, such as Moments in Time~\citep{ref:moments} with 1 million short videos, and Kinetics~\citep{ref:kinetics} with 750k video clips. 
In comparison, medical datasets contain far fewer 3D images. 
For example, Stanford AIMI Coronary Calcium (COCA) dataset~\citep{ref:COCA} only contains 787 CTCAs. 
To generate 3D images, \cite{ref:3dvae} used 991 brain MRI images from the Alzheimer’s Disease Neuroimaging Initiative (ADNI) dataset~\citep{ref:ADNI}. 
Training an effective 3D generative model is difficult with such small datasets of 3D medical images.

Secondly, 3D convolution layers have a large number of parameters, making training time slow and prone to overfitting. 
As a result of 3D convolution, weight parameters take on an additional dimension. 
For example, the conventional $3\times 3$ convolution expands to $3\times 3 \times 3$ in the 3D case. 
Adding the third dimension allows the modeling of the 3D anatomical structure, but it also involves the introduction of an excessive number of parameters and computations, resulting in slower training. 
Moreover, the model is prone to overfitting due to the contrast between the large number of parameters and the small number of 3D training images. 

\begin{figure}[!t]
\centerline{\includegraphics[width=0.45\textwidth]{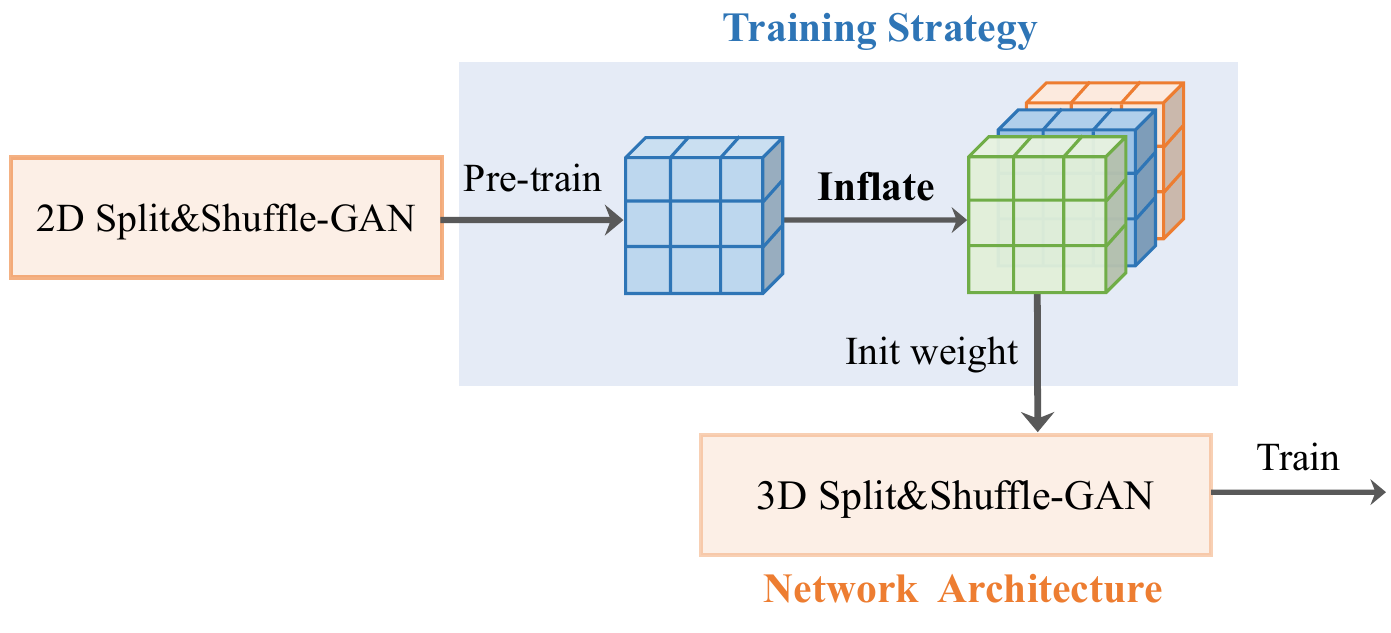}}
\caption{\label{fig:pipeline}The general pipeline of our 3D {generative} model, which includes our contribution to both the training strategy (inflate 2D weights, colored in blue) as well as the network architecture (Split\&Shuffle GAN, colored in orange).}
\end{figure}

To address the above two problems, we propose a novel GAN model, dubbed \textbf{3D Split\&Shuffle-GAN} for effective and efficient 3D medical image generation. 
The proposed model improves existing state-of-the-art GANs (e.g., StyleGAN2~\citep{ref:stylegan2}) from two perspectives: training strategy and network architecture (see Figure~\ref{fig:pipeline}). 

The proposed training strategy takes advantage of the availability of 2D image slices to train a 2D GAN model. It then inflates the 2D weights to initialize the 3D GAN model.
As the 3D GAN model is initialized with informative 2D weights, it can focus more on 3D anatomy, which results in a better generation of 3D images. 
By design, the 2D GAN shares a similar architecture as the 3D GAN, with the exception of the additional convolution dimension (e.g., $3\times 3$ convolution vs. $3\times 3 \times 3$ convolution). 
This enables the 2D weights to be seamlessly expanded to 3D using the weight inflation technique~\citep{ref:inflation}. Since the original inflation was designed for classification models rather than generative models, we evaluate five new inflation variants through extensive experiments to determine the most suitable one for the task of 3D image generation. 

For the network architecture, we devise novel Channel Split\&Shuffle modules to improve both the generator and discriminator networks. 
For the generator, since the state-of-the-art style-based models (e.g., StyleGAN2) incorporate style vectors into convolution weights as a modulated convolution, efficient convolution operations (e.g., depthwise separable convolution~\citep{ref:mobilenet} or group convolution~\citep{ref:alexnet}) cannot be directly adopted. 
This is mitigated by our Split\&Shuffle module, which splits the feature channels into two equal branches and performs a modulated 3D convolution for each branch. 
Then, the output channels are concatenated and shuffled to encourage feature exchanges. With this design, the number of parameters of the generator is reduced by a factor of 2. 
For the discriminator, the number of parameters is further reduced by nearly a factor of 4 by replacing one of the $3\times 3 \times 3$ convolutions with $1\times 1 \times 1$ convolution. 
Although the number of parameters for both the generator and discriminator is significantly reduced, the devised modules achieve a much better performance than the original one. 
Under the extremely data-deficit challenges of generating 3D medical images, our parameter-efficient model is less likely to overfit than the original model. 

To demonstrate the effectiveness of the training strategy and network architecture, we investigated five novel weight inflation variants as well as five network design choices on the heart dataset {(COCA)}. 
In addition, we performed experiments on the brain dataset {(ADNI)} to demonstrate the general applicability of our method. 

To summarize, this paper makes the following contributions to 3D medical image generation:
\begin{itemize}
    \item A novel 3D Split\&Shuffle-GAN model for 3D medical image generation is proposed, and new inflation strategies are developed to facilitate training of 3D medical generation models. 
    \item Parameter-efficient Channel Split\&Shuffle modules are developed for both the generator and discriminator networks, which reduces the number of parameters (by a factor of {at least 2}) and improves generation quality (FID). 
    \item We conducted comprehensive experiments to verify the effectiveness of the inflation strategy and network architecture. We achieved state-of-the-art performance on both the heart and brain datasets. 
\end{itemize}

\section{Related Work}
\label{sec:related}


\subsection{Generative Models for Medical Imaging}
The most popular model for generating synthetic images is the generative adversarial networks (GANs)~\citep{ref:GANs}.
The GANs model synthesizes realistic images from a random noise variable and uses a discriminator to distinguish between the synthesized images and the realistic images. 
The distribution of the synthesized images gradually approaches the distribution of real images with the alternating training of the generator and discriminator. 
State-of-the-art GANs use the style-based generation technique~\citep{ref:stylegan,ref:stylegan2}, in which style vectors are generated (for controlling the style of image generation) from a mapping network. 

Providing annotations to large numbers of images in the field of medical imaging is a challenging task. 
The use of GANs is thus naturally adopted to solve a number of medical problems~\citep{ref:medicalGANs4,ref:medgans2,ref:medgans1,ref:medgans3,ref:MedIAGAN,ref:CondGAN3D,ref:Reviewer2_1,ref:Reviewer2_2,ref:Reviewer2_3,ref:Reviewer2_4}, such as classification, segmentation, registration, low dose CT denoising, {and} MR to {positron emission tomography (PET)} synthesis. 
GANs {were} used by \cite{ref:medicalGANs4} to generate synthetic CT images for data augmentation to enhance liver lesion classification performance. 
RefineGAN~\citep{ref:medgans2} proposes a cyclic consistency loss for the modified variant of the deeper generator and discriminator networks to deal with the {compressed sensing magnetic resonance imaging (CS-MRI)} reconstruction problem. 
To improve the conventional GANs for the low dose CT (LDCT) denoising task, \cite{ref:medgans1} {employed} two practical methods, namely Wasserstein distance and perceptual loss.
A-CycleGAN~\citep{ref:medgans3} makes use of variational autoencoding (VAE), attention, and cycle-consistent generative adversarial network (CycleGAN) to improve existing MR-to-CT image translation algorithms. 
\cite{ref:Reviewer2_2} proposed an effective adversarial U-Net architecture along with different normalization techniques to solve the MRI to PET image synthesis task. 

Despite the wide range of models and {GAN} variants proposed for medical imaging problems, most of them only focus on generating 2D images, disregarding the 3D anatomical structure. 
Only a few attempts have been made to generate 3D images. 
Leveraging an $\alpha$-GAN,  \cite{ref:3dvae} utilizes the variational autoencoder (VAE) and GAN to generate 3D synthetic brain MRI images. 
\cite{ref:3DStylePET} {proposed} a segmentation-guided style-based generative adversarial network (SGSGAN) for synthesizing full-dose PET images, where a style-based generator is directly used for style modulation. 
\cite{ref:HSPN} proposed a hierarchical shape-perception network (HSPN) for 3D brain reconstruction (point cloud) from a single incomplete image. In contrast, our method generates 3D medical images with only random variables as input. 
By extending StyleGAN2's 2D convolutions to 3D convolutions, \cite{ref:3dstylegan} {used} 3D-StyleGAN to generate 3D brain MRI images. 
A comprehensive review of the usage of GANs in 3D data can be found in \cite{ref:3d_review}. 
Since most existing methods lift 2D GANs models to 3D in a straightforward manner, the number of parameters increases significantly, making it challenging to train the model effectively. 
In this paper, we propose both effective training strategies and efficient model architectures to generate 3D medical images using 3D GANs. 

\subsection{Training 3D Convolution Neural Networks}
A multitude of research {effort} has been directed toward 3D CNNs in the field of natural images, especially for the spatiotemporal analysis of videos. 
The main idea is to introduce a third convolution dimension ($k\times k\times k$) to capture the temporal dependencies for video applications such as action recognition~\citep{ref:inflation}. 
The training of 3D CNN models usually relies on large-scale video datasets, e.g., Kinetics-700~\citep{ref:kinetics} with 750k video clips, Moments in Time~\citep{ref:moments} with 1 {million} short videos. 
However, due to the high annotation costs, patient consent issues, and expert annotation challenges, creating 3D medical image datasets of similar scale is not feasible. 
As a result, training 3D medical models is challenging. 

Using degenerated 2D spatial information, another line of work contributes to initializing 3D convolution weights by utilizing beneficial priors. 
For example, \cite{ref:inflation} {proposed} an inflation strategy to stack 2D weights for the 3D weights initialization. The video vision Transformer is trained using the central frame initialization strategy in \cite{ref:vivit}.  
To our knowledge, no similar initialization technique has been explored in 3D medical GANs. On the one hand, the third dimension in video analysis corresponds to temporally varying frames, while the third dimension in medical images describes the 3D anatomical structure. On the other hand, the interplay between the discriminator and generator makes the training process more complex than that of classification models. 
In this paper, we consider both the 3D anatomical structure and the interplay between the discriminator and the generator to facilitate the {3D GAN} training and architecture design. 

\subsection{Parameter-efficient 3D Convolution Neural Networks}
3D convolution neural networks are challenged by the large number of parameters included by the additional third dimension. 
There are two main approaches to addressing this issue: tensor decomposition and efficient module design. 
In tensor decomposition, the low-rank tensor decomposition algorithms are applied to re-calculate the convolution weights, thereby compressing the network and reducing the number of parameters. 
For example, Tensor Train has been used in~\cite{ref:TT}, {CANDECOMP/PARAFAC (CP)} decomposition is applied in~\cite{ref:CP1,ref:CP2}, and Tucker decomposition is adopted in~\cite{ref:tucker}. 
In spite of their mathematical soundness, these methods require specific re-implementation of existing convolution operations and cannot take advantage of the latest hardware acceleration (e.g., the NVIDIA cudnn library). 
In efficient module design, various parameter-efficient modules (e.g., bottleneck~\citep{ref:resnet3d}, group convolution, depthwise separable convolution, and pointwise convolution) are devised to replace the original module. 
These efficient modules are re-arranged and combined to form different network architectures. 
In MobileNet~\citep{ref:mobilenet}, for example, depthwise separable convolutions are used to construct a lightweight deep architecture for mobile devices. 
SqueezeNet~\citep{ref:squeezenet} combines pointwise convolution and regular convolution to form a Fire block. 
The computation cost of ShuffleNet~\citep{ref:shufflenet} is reduced by using pointwise group convolution.  
A comprehensive analysis of these modules can be found in \cite{ref:resource}. 

All the above parameter-efficient designs are based on classification models, which cannot be directly and easily adopted in 3D generative models, such as StyleGAN2. 
StyleGAN2's style modulation mechanism will be destroyed if these modules are trivially adopted. 
To address this issue, we propose customized 3D modules for the style-based generative models to enable parameter-efficient generation of 3D medical images. 


\section{Methods}
\label{sec:method}

\begin{figure*}[ht]
\centerline{\includegraphics[width=0.73\textwidth]{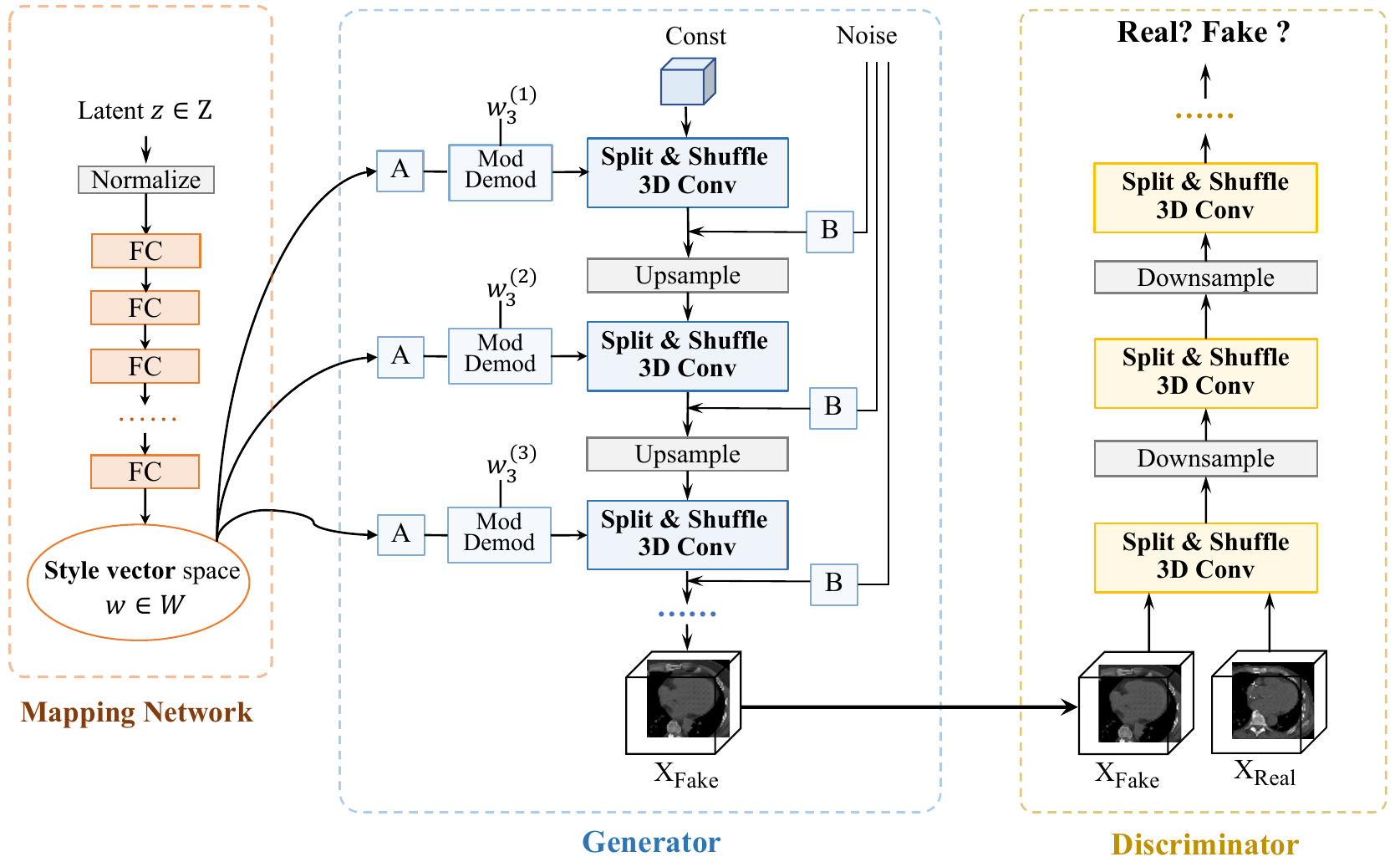}}
\caption{\label{fig:architecture}The overall architecture of the proposed 3D Split\&Shuffle-GAN. It is composed of a Mapping Network, a Generator, and a Discriminator. Mapping Network maps a latent variable $z$ to the style vector space $\mathcal{W}$ and produces the per-layer style vectors with a learned affine transform (A). The generator starts from a constant 3D input. Then it controls the 3D generation styles with the per-layer style vectors and adds details from the per-channel scaled (B) noise input. Weight Mod\&Demod incorporates the style vector into the convolution operation ($w_3^{(1)}, w_3^{(2)}, w_3^{(3)}, \dots$ represent the 3D convolution weights for first, second, third, \dots layers). Discriminator tries to differentiate the real 3D images from the generated fake 3D images. Inside both the Generator and Discriminator, we devise novel \emph{\textbf{Channel Split$\&$Shuffle}} modules for parameter-efficient 3D convolution {operations}, which are customized for the style-based generation framework.}
\end{figure*}

\subsection{Preliminary of 3D Medical Image Generation}
\label{sec:stylegan2}


\subsubsection{Overview of StyleGAN2 model} 

\paragraph{Mapping Network} 
A key difference between style-based {generative} models (e.g., StyleGAN2) and previous GANs is the introduction of the mapping network $f$. 
Specifically, given a latent code $z \in \mathcal{Z}$, $f: \mathcal{Z}\rightarrow \mathcal{W}$ first produces a vector $w \in \mathcal{W}$. The learned affine transform A is then applied to $w$ to obtain the generator's per-layer style vectors $s$. 

\paragraph{Generator} 
In the generator, original StyleGAN~\citep{ref:stylegan} directly utilizes the style vectors for adaptive instance normalization (AdaIN) on the feature maps, which will cause characteristic artifacts such as droplets. To mitigate these unrealistic artifacts, StyleGAN2 incorporates the style vectors into the weight modulation (Mod) operation, then applies the demodulation (Demod) to serve as the instance normalization. 
\begin{align}
    \label{eq:modulation}
    \text{Modulation: }   & w^{'}_{ijk} = s_i \cdot w_{ijk}\,,\\
    \label{eq:demodulation}
    \text{Demodulation: } & w^{''}_{ijk} = w^{'}_{ijk} / \sqrt{\sum_{i,k} {w^{'}_{i,j,k}}^2 + \epsilon} \,,
\end{align}
where $w, w^{'}, w^{''}$ are the original, modulated and demodulated convolution weights, $s_i$ is the style vector corresponding to the $i$-th feature map, $j,k$ iterate the output feature maps and the spatial resolution, $\epsilon$ is a small constant to avoid numerical issues. 
In the above modulation and demodulation operations, the style vectors are directly entangled with convolution weights, which removes the characteristic artifacts while retaining the style controllability. 
However, this also impedes the straightforward modification to the convolution layers, e.g., depthwise separable convolution (details in Sec.~\ref{sec:architecture}). 

\paragraph{Discriminator} 
The discriminator of StyleGAN2 introduces a minibatch standard deviation layer to calculate the deviation of a minibatch and concatenates it to the original feature maps. This reduces the dependency on a minibatch to encourage diverse generations. 


\subsubsection{3D Medical Image Generation} 
The StyleGAN2 {was} originally designed for 2D natural image generation. 
To apply it {to} the {generation of} 3D medical images, a straightforward approach~\citep{ref:3dstylegan} is by lifting all the 2D convolution operations to the 3D convolution operations, e.g., expanding the $3\times 3$ convolutions to the $3\times 3\times 3$ convolutions. 
Albeit simple, this approach will significantly increase the number of parameters, thereby posing two practical issues: (1) it will require a large number of 3D images for training; otherwise, the model suffers overfitting and mode collapse issues\footnote{Overfitting means that on small datasets, the discriminator overfits to the training examples and training starts to diverge~\citep{ref:ADA}. Mode collapse means that the generator only produces a limited set of outputs.}, {and} (2) the largely increased parameter number will slow down training and generation. 
However, no existing methods have simultaneously addressed both issues. 
Therefore, it is non-trivial to improve StyleGAN2 for 3D medical image generation. 


In this paper, we deal with the issues from the \textbf{training strategy} (Weight Inflation in {Section}~\ref{sec:inflation}) and \textbf{network architecture} (Split\&Shuffle in {Section}~\ref{sec:architecture}) perspectives and propose an efficient 3D generative model as shown in Figure~\ref{fig:architecture}. 



\subsection{Inflating 2D Convolution Weights} 
\label{sec:inflation}
This section discusses how to design a training strategy to generate synthetic 3D medical images and overcome the issue of data scarcity. 
Transfer learning~\citep{ref:transfer} is a widely-used technique to overcome the data paucity of the target task by employing additional external datasets. 
As an example, ImageNet~\citep{ref:imagenet} is used as an external dataset for a variety of tasks, including object detection, and segmentation. 
In MinGAN~\citep{ref:transferGAN}, knowledge is transferred from GANs to domains with few images. 
It is currently difficult to apply the transfer learning technique for 3D {GAN} models 
in medical research due to the lack of large datasets and effective transfer learning strategies. 
3DSeg-8 dataset~\citep{ref:med3d} has been aggregated from eight datasets to facilitate transfer learning between 3D medical images for liver segmentation and nodule classification. 
However, there are only tens to hundreds of 3D images of organs/tissues in each dataset in 3DSeg-8, which makes GANs incapable of transferring detailed knowledge. 
Moreover, the interplay between the generator and discriminator poses a different transfer learning challenge compared to classification tasks with a single network. 

Although direct transferring from external datasets is {not feasible} for 3D medical image generation, we note an interesting and helpful observation: \emph{\textbf{2D slices in a medical dataset are several magnitudes larger than 3D images}}. 
As an example,  COCA~\citep{ref:COCA} contains only 787 CTCA images, but these CTCA images contain $39,281$ 2D slices. 
{Hence}, the number of these 2D slices is sufficient to train a 2D generative model such as StyleGAN2. 
Since 2D and 3D generative models have distinct weights, it is not possible to transfer weights directly from 2D to 3D generative models. 
As a result, we are naturally drawn to another technique called weight inflation~\citep{ref:inflation}, which enables effective 3D network training from the pre-trained 2D weights. 

Weight inflation was first introduced in~\cite{ref:inflation} for the design and training of 3D video action recognition networks. 
The method has been applied to both CNNs and Transformers-based 3D models~\citep{ref:inflation,ref:vivit,ref:transfer3D}. 
In technical terms, it extends/inflates/copies the 2D convolution weights along the third dimension (e.g., temporal dimension in the video) to provide a more favorable initialization for the 3D convolution networks.
To ensure feasible inflation from 2D weights to 3D models, the 2D and 3D networks must have the same basic structure except for the additional third dimension (e.g., the $3\times 3$ and $3\times 3\times 3$ convolutions should have the same number of channels). 
\emph{\textbf{To our knowledge, this technique has not been applied to medical image analysis for effective 3D medical image generation}}. 

Considering the application differences between video action recognition and 3D medical image generation, we propose five customized inflation strategies to facilitate the training of 3D StyleGAN2. 
Here, we set the size of convolution weights to be $3\times 3\times 3$, but our strategies can be easily adapted to other weight sizes. 
Let $w_2 \in \mathbb{R}^{C_I\times C_O\times 3\times 3}$ denote the pre-trained 2D convolution weights and $w_3 \in \mathbb{R}^{C_I\times C_O\times 3\times 3\times 3}$ denote the corresponding 3D convolution weights ($C_I$ represents the number of input channels, $C_O$ represents the number of output channels). At first, we initialize $w_3$ from a random Gaussian $\mathcal{N}(0, 0.1)$. 
Then, the weight $w_3$ is modified by the following inflation strategies:
\begin{itemize}
\item \textbf{Inflate-1}: only inflating 1 center dimension. 
\begin{equation}
    w_3[:,:,1,:,:] = w_2\,.    
\end{equation}

\item \textbf{Inflate-2}: inflating 2 dimensions. 
\begin{equation}
    w_3[:,:,0,:,:] = w_2,\quad w_3[:,:,1,:,:] = w_2\,.   
\end{equation}
\item \textbf{Inflate-3}: inflating all 3 dimensions.
\begin{align}
    \nonumber
    & w_3[:,:,0,:,:] = w_2,\quad w_3[:,:,1,:,:] = w_2, \\
    & w_3[:,:,2,:,:] = w_2\,. 
\end{align}
\item \textbf{Inflate-ASC}: inflating the axial, sagittal, and coronal planes of 3D medical views. 
\begin{align}
    \nonumber
    & w_3[:,:,1,:,:] = w_2,\quad w_3[:,:,:,1,:] = w_2, \\
    & w_3[:,:,:,:,1] = w_2\,. 
\end{align}
\item \textbf{Inflate-NWI}: the negative weight initialization (NWI) is used here, with the center dimension owing a larger value and the other dimensions owing negative values ($T=3$). 
\begin{align}
    & w_3[:,:,i,:,:] = \alpha_i * w_2, \; \alpha_i = 
    \left\{
        \begin{array}{ll}
      \frac{2T-1}{T}, \text{if } i=1 \\
      -\frac{1}{T}, \text{otherwise}\,.\\
    \end{array} 
    \right. 
\end{align}
\end{itemize}

By design, different inflation strategies offer diverse ways of reusing the 2D weights: the reusing degree increases from Inflate-1 to Inflate-3; Inflate-ASC considers the anatomical views; Inflate-NWI modifies Inflate-1 with more attention on the center dimension and negative weights on others. 
Intuitively, these inflation strategies introduce helpful 2D structure priors through weight initialization, which significantly reduces the training burden of the 3D convolution weights. 
In this way, by focusing more on the third dimension (for anatomy learning), the generative model is able to generate high-quality 3D images quickly and efficiently.




\begin{figure*}[ht]
\centerline{\includegraphics[width=0.72\textwidth]{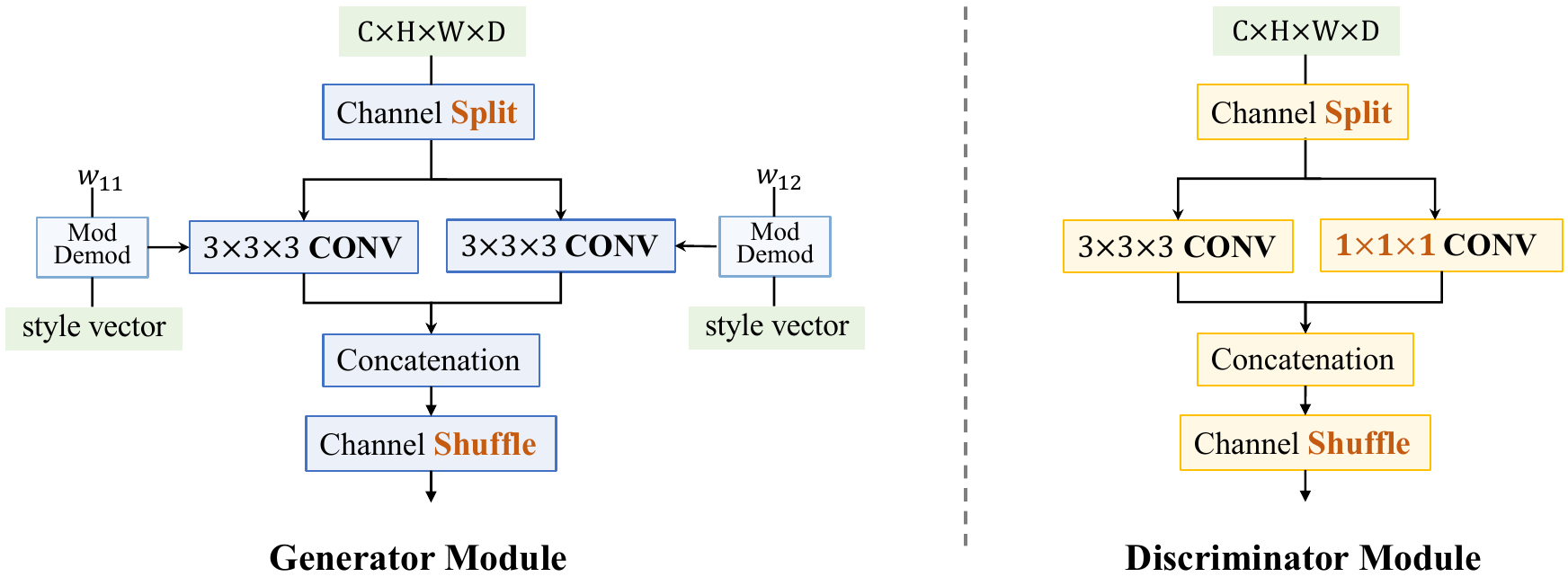}}
\caption{\label{fig:module}The proposed Channel Split\&Shuffle Convolution Modules for the Generator and Discriminator. The Discriminator module has two differences from the Generator module: the style Mod\&Demod and $1\times 1\times 1$ Convolution. Overall, the proposed module reduces the number of parameters by a factor of 2 in {the} Generator and nearly by a factor of 4 in the Discriminator. Here, $w_{11}$ and $w_{12}$ denote the 3D convolution weights of the left and right branches, which are used to perform {the} Mod\&Demod operation in {Equations}~\ref{eq:modulation}-\ref{eq:demodulation}.} 
\end{figure*}

\subsection{Efficient 3D Architecture Design}
\label{sec:architecture}

Although inflation strategy mitigates the lack of 3D data in training the 3D GAN model, it still suffers from the large number of model parameters. 
In this section, we address this issue from the perspective of efficient 3D architecture design. 

We observe that most of the parameters in the 3D neural network architecture originate from the 3D convolution operation, which extends the 2D convolution weights to 3D, to model the 3D contextual and anatomical structures (e.g., lifting $3\times 3$ weights to $3\times 3\times 3$). 
Existing efficient 3D architecture designs mainly focus on parameter-efficient 3D convolution. 
On the one hand, factorized high-order CNNs are proposed with different tensor decomposition algorithms such as Tensor-Train (TT)~\citep{ref:TT}, CP decomposition~\citep{ref:CP1,ref:CP2}, Tucker Decomposition~\citep{ref:tucker}. 
These methods compress networks and reduce their parameters by applying low-rank tensor decompositions to high-order weights. 
On the other hand, driven by the requirements of mobile devices, various parameter-efficient convolution variants {have been } devised and combined into efficient architectures, such as group convolution~\citep{ref:alexnet}, bottleneck~\citep{ref:resnet}, depthwise separable convolution~\citep{ref:mobilenet}. 
Group convolution divides the channels into groups and performs convolution only within each group. Bottleneck was introduced in ResNet~\citep{ref:resnet} to reduce the number of channels of the $3\times 3$ convolution by wrapping it with two $1\times 1$ convolutions. 
With depthwise separable convolutions, the standard convolutions are factorized into a depth-wise convolution (i.e., a group convolution with a group number equal to a channel number) followed by a $1\times 1$ pointwise convolution. 


The above designs were developed to improve the efficiency of various applications, e.g., HO-CPConv~\citep{ref:CP2} for spatiotemporal facial emotion analysis, MobileNet~\citep{ref:mobilenet} for image classification and object detection, and 3D-MobileNet~\citep{ref:resource} for video action recognition. 
\emph{\textbf{Despite this, the efficient 3D GANs architecture is rarely studied, especially for the state-of-the-art StyleGAN2 model.}} 
We attribute this to two possible reasons: (1) The StyleGAN2 architecture is more delicate than classification models, hindering the straightforward adoption of existing modules such as tensor decomposition, group convolution, or depthwise separable convolution. Specifically, the style vectors are absorbed in the modulation and demodulation operations ({Equations}~\ref{eq:modulation}, \ref{eq:demodulation}), which sets up hurdles for existing modules. 
(2) As a result of the interplay between the discriminator and generator, GANs training is difficult. 
It is non-trivial to directly use the same modules for the discriminator and generator to achieve the best performance. 

Based on the above analysis, we propose a unique design customized to the StyleGAN2 model for parameter-efficient generation (Figure~\ref{fig:module}). 
For the generator, since the direct adoption of existing efficient modules (e.g., group convolution, depthwise separable convolution) will break the entangled structure of the convolution weights and style vectors ({Equations}~\ref{eq:modulation}, \ref{eq:demodulation}), we equally \textbf{Split} the feature maps, using the channel split operation, to {create} two branches. 
The modulation and demodulation operations {for} the style vectors and convolution weights are individually applied to each branch. 
{Afterwards}, the outputs of the two branches are concatenated, and the \textbf{Channel Shuffle} operation is performed, allowing information to be shared between two channels. 
If the Channel Shuffle operation is not performed, the generator is considered to be two independent networks. 
Channel Shuffle enables hybrid and diverse pattern combinations across branches to facilitate image generation quality. 
For the discriminator, as neither modulation nor demodulation is applied, there is more flexibility in improving the design. Therefore, we {devised} two asymmetric branches with $3\times 3\times 3$ and $1\times 1\times 1$ convolutions. The $1\times 1\times 1$ convolution leads to a further parameter reduction while compromising the local spatial structure. But this is rectified by the Channel Shuffle operation, which exchanges information by shuffling the feature maps of $1\times 1\times 1$ and $3\times 3\times 3$ convolutions. 

Considering $C=32$ in Figure~\ref{fig:module}, both the input and {output} feature maps are of size $32\times H\times W\times D$. Without splitting, the size of the convolution weight is $32\times 32 \times 3 \times 3 \times 3$. 
With splitting, the input map is split into two $16\times H\times W\times D$ branches, each undergoing a 3D convolution (weight size $16\times 16\times 3\times 3\times 3$). The outputs of two branches are concatenated to get the feature map of size $32\times H\times W\times D$. 
{So,} the total size of the convolution weight is $2\times 16\times 16\times 3\times 3\times 3$, which means the generator enjoys a parameter reduction of 2. 
Similarly, the discriminator has a total weight size of $16\times 16\times 3\times 3\times 3 + 16\times 16\times 1\times 1\times 1$, which means it enjoys a parameter reduction of nearly 4 (3.857). 


We also propose several other parameter-efficient convolution architectures as baselines to verify the Split\&Shuffle design's effectiveness. 
As a result of the modulation and demodulation constraint in the generator, several baselines only modify the discriminator (i.e., D only). 
All the model variants are listed below:
\begin{itemize}
    \item \textbf{Group Convolution (D only)}, which replaces the convolution in the discriminator with a group convolution. 
    \item \textbf{Depthwise Separable Convolution (D only)}, which replaces the convolution in the discriminator with a depthwise separable convolution. 
    \item \textbf{Split\&Shuffle Convolution (D only)}, which applies the Split\&Shuffle module only on the discriminator. 
    \item \textbf{Split Convolution}, which applies the channel split without channel shuffle. 
    \item \textbf{Split\&Shuffle Convolution}, which is our final design. 
\end{itemize}

\section{Results}
\label{sec:experiment}

\subsection{Datasets and Evaluation Metrics}
\subsubsection{Datasets} Stanford AIMI Coronary Calcium (COCA)~\citep{ref:COCA} dataset is used for heart CTCA images. 
COCA contains 787 3D coronary CT images. 
Each 3D image has a different number (ranging from $27$ to $156$) of 2D slices on the axial plane, and they add up to $39,281$ axial slices in total. 
For brain MRI images, we {used} the Alzheimer’s Disease Neuroimaging Initiative (ADNI) {dataset}~\citep{ref:ADNI}. Specifically, we {used} $991$ T1 structural images from the Cognitively Normal (CN) research group. 
MR images {from} non-brain areas {were} removed by the dataset provider using the software FreeSurfer's\footnote{https://surfer.nmr.mgh.harvard.edu/fswiki} recon-all function. 
The processed MR images have $256$ slices from all three planes. 

\subsubsection{Evaluation Metric} 
In order to assess the quality of generated images, GANs usually use the Fréchet inception distance (FID)~\citep{ref:fid} metric, which compares the feature distributions of real and generated images. 
By default, the Inception V3 network~\citep{ref:inceptionV3} pre-trained with 2D natural images was deployed for feature extraction. 
However, the method cannot be directly applied to 3D images. 
As such, taking into account the 3D medical structure, we {measured} the FID scores on the center slices of axial, sagittal and coronal planes, i.e., \emph{FID-ax}, \emph{FID-sag}, and \emph{FID-cor}. 
Lastly, we {averaged} the three FID scores to obtain \emph{FID-avg} as an overall measurement. 

However, the FID alone cannot evaluate 3D medical image generation comprehensively. 
The reasons are {twofold}: (1) the inception model only accepts 2D image slices of a specific plan, and (2) the inception model is  pre-trained on natural images with a large gap with medical images. 
Therefore, we also {adopted} the widely-used metrics MS-SSIM, PSNR, and the t-Distributed Stochastic Neighbour Embedding (t-SNE) to evaluate the performance. 



\subsection{Implementation Details}
\subsubsection{Pre-processing} 
Both the COCA and ADNI images {were} resized to $64\times 64$ and $128\times 128$ using bilinear interpolation in the sagittal and coronal planes. 
We {aligned} the axial slice number of the COCA dataset to $32$ either by consecutive slice sampling or by zero padding. 
For the ADNI dataset, we {resized} the axial slice number to $64$ {directly}. As a result, the image resolution of COCA and ADNI datasets are $32\times 64\times 64$ ($32\times 128\times 128$) and $64\times 64\times 64$, respectively. 
The Hounsfield Unit values {were} clipped to $[-250, 650]$ and then normalized to $[-1, 1]$.

\subsubsection{StyleGAN2 architecture} 
The base layer (Const in Figure~\ref{fig:architecture}) for the COCA dataset {was} set to $2\times 4\times 4$, followed by $5$ upsampling stages. 
For the ADNI dataset, the base layer {was} set to $4\times 4\times 4$, followed by $5$ upsampling stages. The total number of convolution channels {was} set to $32$, with each branch owing $16$. 
The feature dimension of the mapping network {was} set to $64$. 

\subsubsection{Hyper-parameters}
We {followed} the StyleGAN2 configuration for training, with the following exceptions: $\gamma=0.0512$ for $R_1$ regularization, $\text{minibatch}=128$, learning rate $\text{lr}=0.0025$ for training from scratch, $\text{lr}=0.002$ for inflation initialization. For 2D GANs pre-training, the models are fed with $25,000$K images in total. {However, the 3D GANs were} trained with $5,000$K images since 3D models are slow to train.


\begin{figure*}[ht]
     \centering
     \begin{subfigure}[b]{0.32\textwidth}
         \centering
         \includegraphics[width=\textwidth]{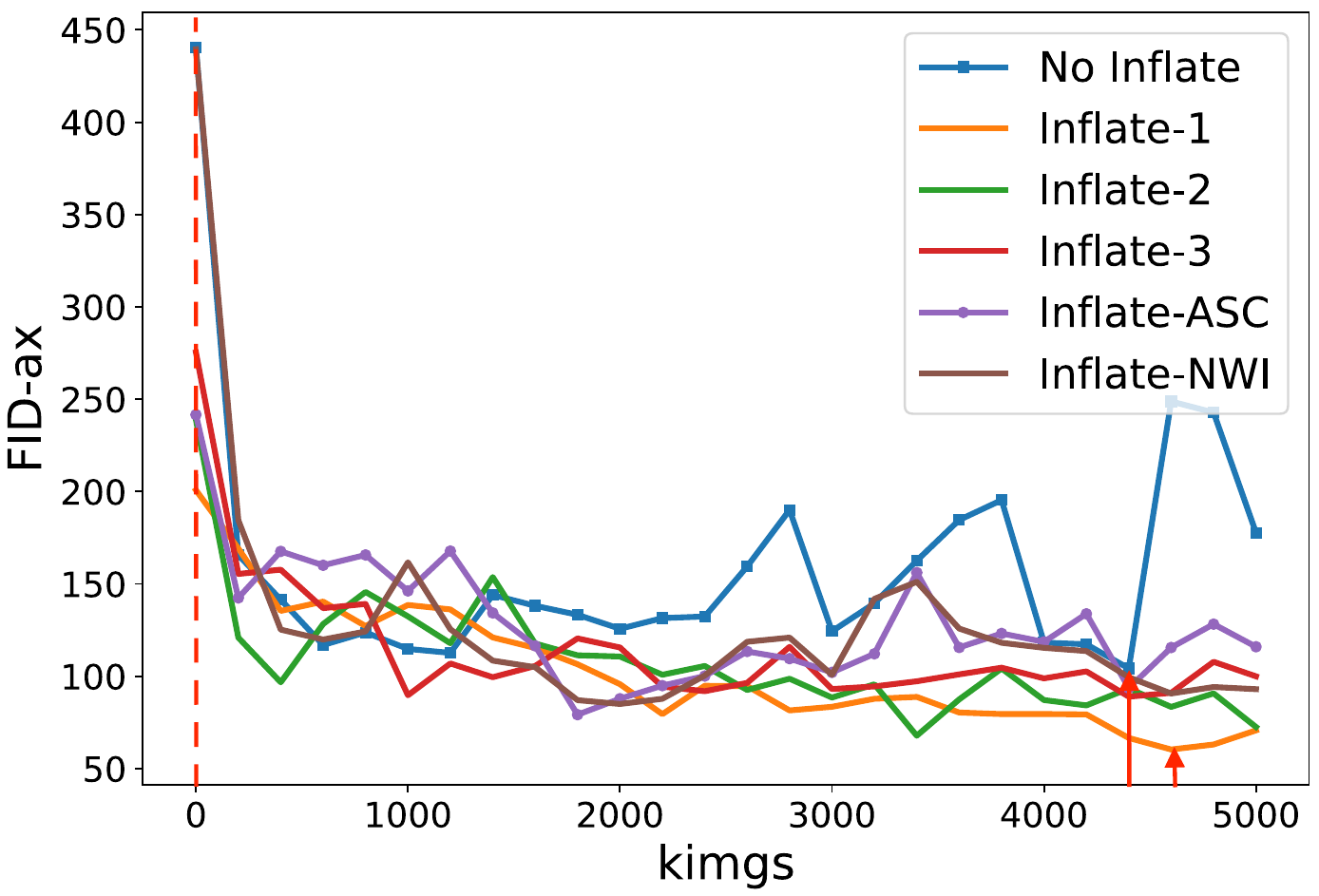}
         \caption{\label{fig:infAx}Axial}
     \end{subfigure}
     \hfill
     \begin{subfigure}[b]{0.32\textwidth}
         \centering
         \includegraphics[width=\textwidth]{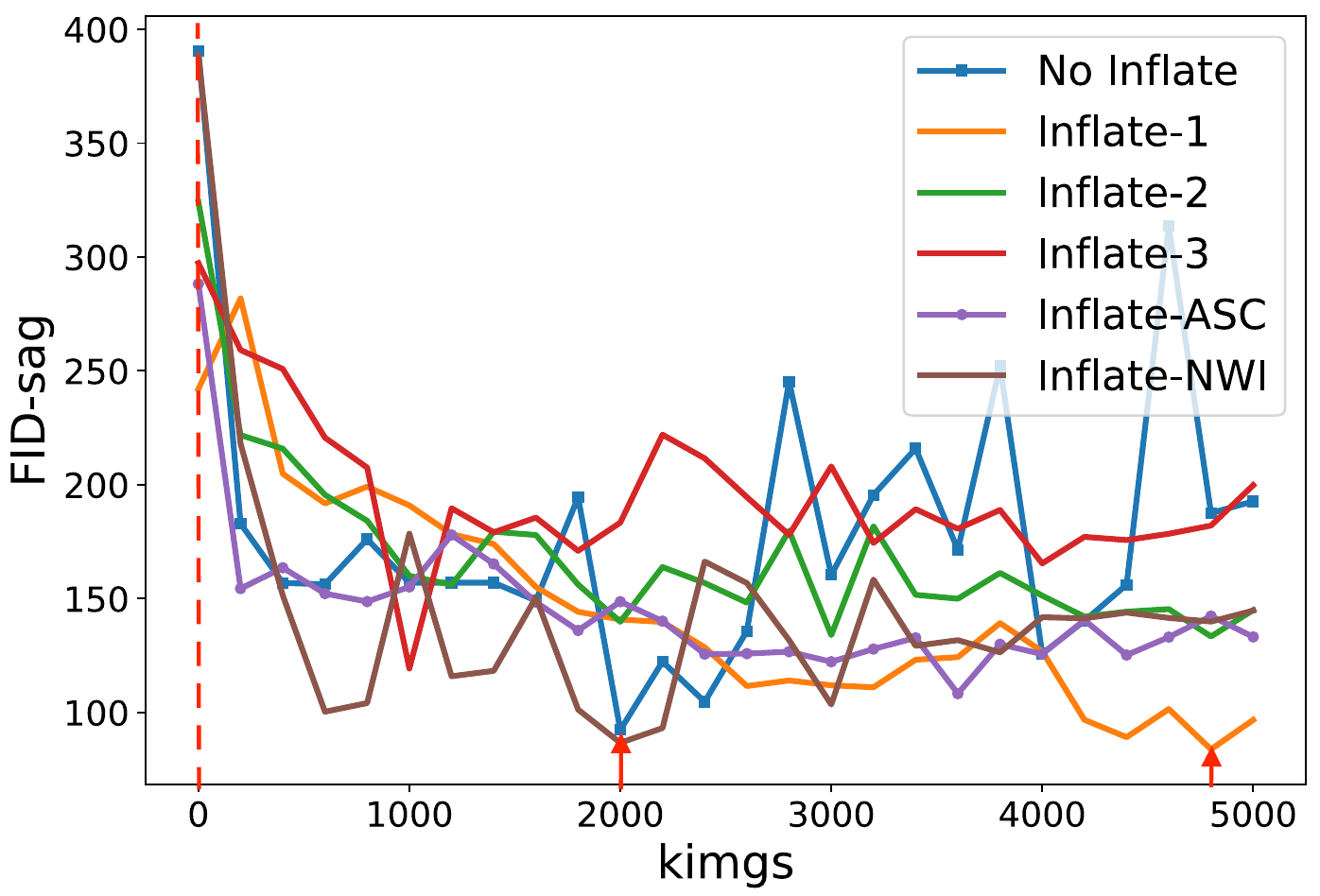}
         \caption{\label{fig:infSag}Sagittal}
     \end{subfigure}
     \hfill
     \begin{subfigure}[b]{0.32\textwidth}
         \centering
         \includegraphics[width=\textwidth]{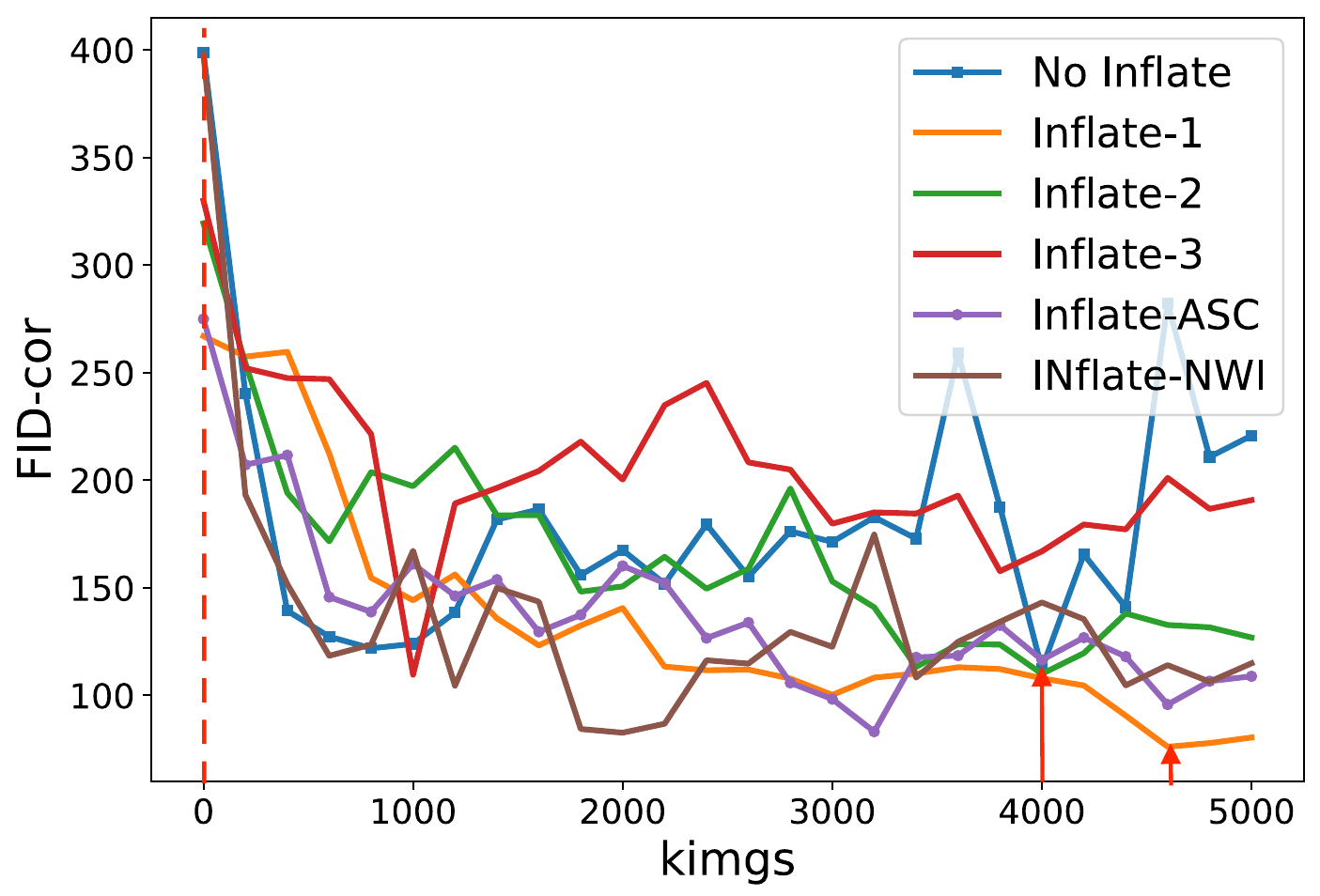}
         \caption{\label{fig:infCor}Coronal}
     \end{subfigure}
    \caption{\label{fig:inflation1}Performance of different inflation variants during the training process (kimgs). Among all, ``Inflate-1'' generally obtains good FID scores on all three views. Red arrows show the best FID iterations (kimgs) for ``No Inflate'' and ``Inflate-1''.} 
\end{figure*}

\subsection{Inflation and Architecture Analysis} 
\subsubsection{Inflation} 
The first step is to conduct comprehensive experiments to verify that inflation strategies are effective for initializing the 3D generative model. 
A 2D StyleGAN2 model is pre-trained using all the $39,281$ axial slices to obtain the 2-dimensional convolution weights. Since the number of images is sufficient, the 2D model achieves an FID of 7.71. This means that the pre-trained 2D weights capture rich slice-level contextual information to generate high-quality 2D slices. 
Thus, it verifies the rationale behind inflating 2D pre-trained weights for 3D generative models.

Starting from the same pre-trained 2D weights, we apply all the proposed inflation strategies as well as a ``No Inflate'' baseline to initialize and train the 3D StyleGAN2 model. 
The results are shown in Table~\ref{tab:inflation}. 
It can be seen that three inflation strategies outperform the ``No Inflate'' baseline (the other two are comparable), indicating that inflation strategies are generally effective as favorable initialization methods for 3D generative models. 
``Inflate-1'', which only initializes one center dimension, achieves the best performance (FID-avg) among all inflation variants. 
Performance gradually degrades as more weights are initialized from 2D weights (``Inflate-2'' and ``Inflate-3''). 
We {hypothesise} that the overly inflated 3D weights will prevent the model from freely learning the third-dimensional anatomical structure. 
``Inflate-ASC'' and ``Inflate-NWI'' perform slightly worse than ``Inflate-1''.

\begin{table}[t]
\centering
\caption{\label{tab:inflation}Performance of different inflation variants {(lower numbers are better)}.}
\setlength{\tabcolsep}{8pt}
\resizebox{0.4\textwidth}{!}{
\begin{tabular}{llll|l}
\hline
Model       & FID-ax & FID-sag & FID-cor & FID-avg \\ \hline
No Inflate  & 104     & 92      & 112      &  102.7   \\
Inflate-1   & \textbf{60}      & \textbf{83}      & \textbf{76} & \textbf{73.0}    \\
Inflate-2   & 68     & 133      & 110      & 103.7  \\
Inflate-3   & 89     & 119      & 110      & 106.0   \\
Inflate-ASC & 79     & 108      & 83       & 90.0   \\
Inflate-NWI & 85     & 87       & 83       &  85.0\\\hline 
\end{tabular}
}
\end{table}

\begin{figure}[h]
     \centering
     \begin{subfigure}[b]{0.45\textwidth}
         \centering
         \includegraphics[width=\textwidth]{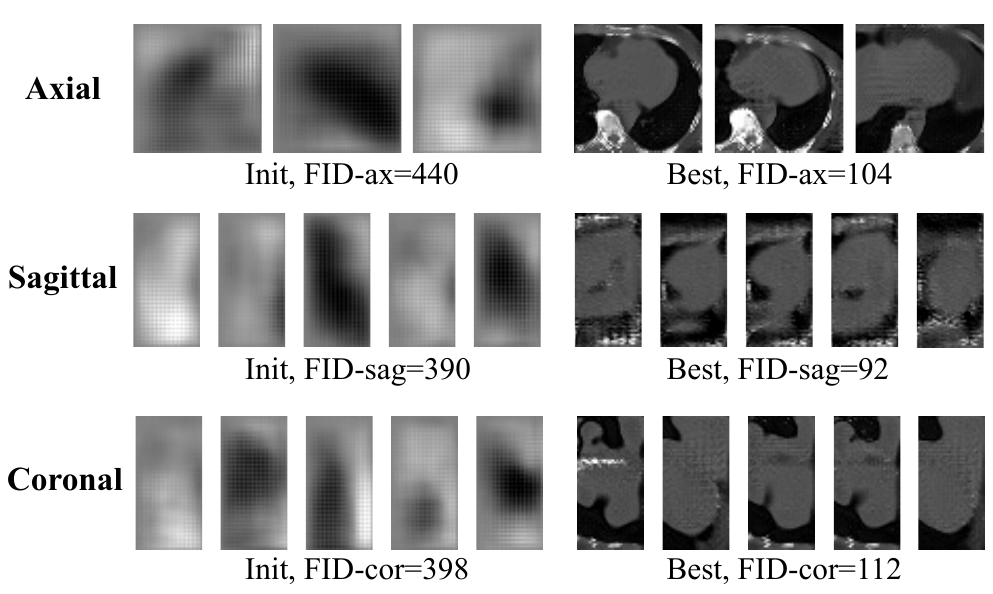}
         \caption{\label{fig:inf_baeline}No Inflate}
     \end{subfigure}
     \hfill
     \begin{subfigure}[b]{0.45\textwidth}
         \centering
         \includegraphics[width=\textwidth]{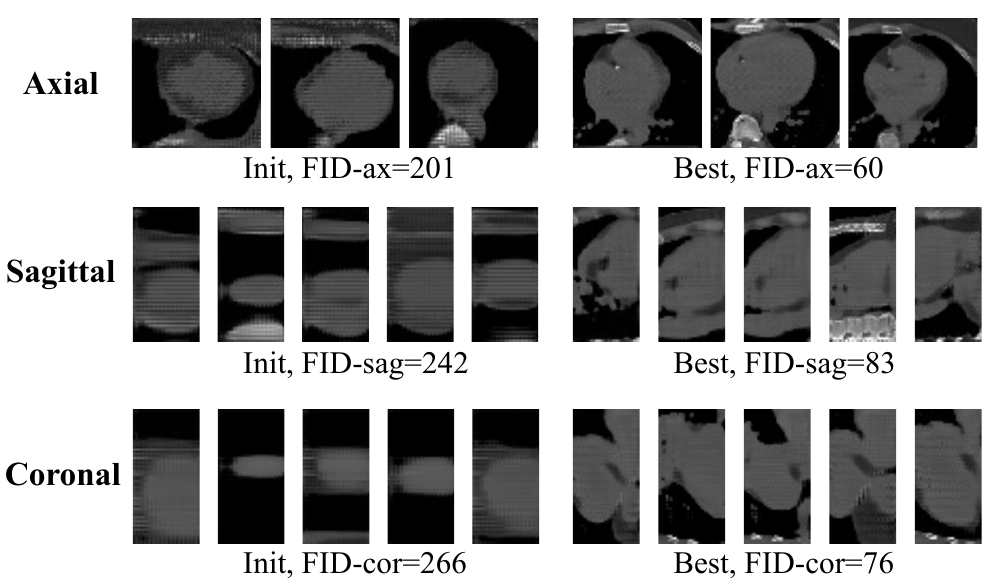}
         \caption{\label{fig:inf_inflate}Inflate-1}
     \end{subfigure}
    \caption{\label{fig:inflation2}Generated images from the initial training (dashed lines in Fig.~\ref{fig:inflation1}) and the best (red arrows in Fig.~\ref{fig:inflation1}) training iterations. (a) With random weights as an initialization (No Inflate), the initially generated images are meaningless.  (b) With inflated weights as a favorable initialization (Inflate-1), the initial generated images already show basic {anatomical} structures.} 
\end{figure}


To  understand the training dynamics of various inflation strategies, we plot the FID scores with respect to the training iteration (measured by the number of $1,000$ training images, i.e., kimgs) in Figure~\ref{fig:inflation1}. 
In general, ``Inflate-1'' achieves the best performance during training, consistent with Table~\ref{tab:inflation}. 
Most inflation variants achieve significantly better FID scores than the baseline at the start of training (i.e., kimgs=0), demonstrating the effectiveness of 2D {weight} priors in training StyleGAN2. 
``Inflate-NWI'' initially performs worse because it modifies the original 2D weights, but in the end, it outperforms the baseline since the prior informative 2D weights have made a big difference.

\begin{table}[t]
\centering
\caption{\label{tab:inflationGD}Inflation on Generator (G) and Discriminator (D).} 
\setlength{\tabcolsep}{8pt}
\resizebox{0.45\textwidth}{!}{
\begin{tabular}{lllll|l}
\hline
Model                    & Module & FID-ax & FID-sag & FID-cor & FID-avg \\ \hline
\multirow{3}{*}{Inflate-1}   & G    & 105 & 82 & 88 & 91.7   \\
                             & D    & 85  & 94 & 104 & 94.3    \\
                             & G\&D & 60  & 83 & 76 & \textbf{73.0}    \\\hline
\multirow{3}{*}{Inflate-ASC} & G    & 93 & 96 & 94 & 94.3  \\ 
                             & D    & 123& 113& 127& 121.0  \\ 
                             &G\&D  & 79 & 108& 83 & \textbf{90.0}  \\ \hline
\multirow{3}{*}{Inflate-NWI} & G    & 85 & 91 & 94 & 90.0   \\
                             & D    & 90 & 92 & 98 & 93.3\\
                             & G\&D & 85 & 87 & 83 & \textbf{85.0} \\
\hline 
\end{tabular}
}
\end{table}

In Figure~\ref{fig:inflation2}, we randomly {generated} the axial/sagittal/coronal slices of CT images for both the ``No Inflate'' baseline and our best variant ``Inflate-1'' to intuitively investigate why the inflation strategy works. Specifically, we show the generated images from the initial training iteration (i.e., the dashed lines in Figure~\ref{fig:inflation1}) and the best training iteration (i.e., the red arrows in Figure~\ref{fig:inflation1}). 
It is easy to observe that with inflation as a favorable initialization, the generated images already show meaningful anatomical structures even before training (e.g., FID-ax=201). 
{In} contrast, the randomly initialized ``No Inflate'' generates blurry meaningless images before training (FID-ax=440). 
This comparison provides an intuitive explanation for the working mechanism of the inflation strategy: with effective inflation, the 3D generative model can inherit meaningful 2D anatomical priors for better subsequent training. 
Furthermore, starting from better initial weights, the inflated model (``Inflate-1'') is trained to achieve superior {generative} performance compared to the ``No Inflate'' baseline (e.g., FID-ax=60 vs FID-ax=104).

StyleGAN2's discriminator and generator have different structures by design. 
This motivates us to examine how inflation affects the discriminator and the generator. 
Specifically, we {selected} the three best inflation variants (``Inflate-1'', ``Inflate-ASC'', and ``Inflate-NWI'') and {performed} three sets of experiments (Table~\ref{tab:inflationGD}): inflating the generator only (G), inflating the discriminator only (D), and inflating both the generator and discriminator (G\&D, default). 
This analysis reveals two observations: (1) inflating the entire model (G\&D) always achieves the best performance, (2) the generator plays a more important role in the inflation strategy, which is reasonable because the generator is responsible for generating images with the style vectors to control the generation.



\begin{table}[ht]
\centering
\caption{\label{tab:architecture}Parameter and performance comparison on different architectures. Models with ``-D" mean only the discriminator owns the modified architecture.}
\setlength{\tabcolsep}{5pt}
\resizebox{0.45\textwidth}{!}{
\begin{tabular}{ll|lll|l}
\hline
Model                 & \#param & FID-ax & FID-sag & FID-cor & FID-avg \\ \hline
Baseline              & 0.600M          & 104 & \textbf{92}  & 112 & 102.7    \\
Group-D               & 0.434M          & 80  & 130 & 105 & 105.0 \\
Depthwise-D           & 0.367M          & \textbf{67}  & 136 & 117 & 106.7  \\
Split\&Shuffle-D        & 0.416M          & 84  & 121 & 118 & 107.7  \\
Split                 & \textbf{0.291M} & 109 & 151 & 146 & 135.3   \\
Split\&Shuffle          & \textbf{0.291M} & 85  & 113 & \textbf{91}  & \textbf{96.3}   \\ \hline 
\end{tabular}
}
\end{table}

\begin{table}[ht]
\centering
\caption{\label{tab:infArc}Model performance when inflation strategy (``Inflate-1'') is applied to different architectures. FID\_2D denotes the pre-trained 2D model performance.}
\setlength{\tabcolsep}{4pt}
\resizebox{0.45\textwidth}{!}{
\begin{tabular}{l|ll|lll|l}
\hline
Model                 & \#param & FID-2D & FID-ax & FID-sag & FID-cor & FID-avg \\ \hline
Baseline              & 0.600M & \textbf{7.7} & 60  & 83 & 76  & 73.0 \\
Group-D               & 0.434M & 10.9         & 78  & 76  & 76  & 76.7 \\
Depthwise-D           & 0.367M & 14.0         & 87  & 129 & 84  & 100.0 \\
Split\&Shuffle-D        & 0.416M & 10.6         & 95  & 112 & 101 & 102.7 \\
Split                 & \textbf{0.291M} & 12.2& 120 & 144 & 106 & 123.3\\
Split\&Shuffle          & \textbf{0.291M} & 11.6& \textbf{46} & \textbf{66} & \textbf{63}   & \textbf{58.3} \\
\hline 
\end{tabular}
}
\end{table}

\begin{figure}[h]
\centerline{\includegraphics[width=0.48\textwidth]{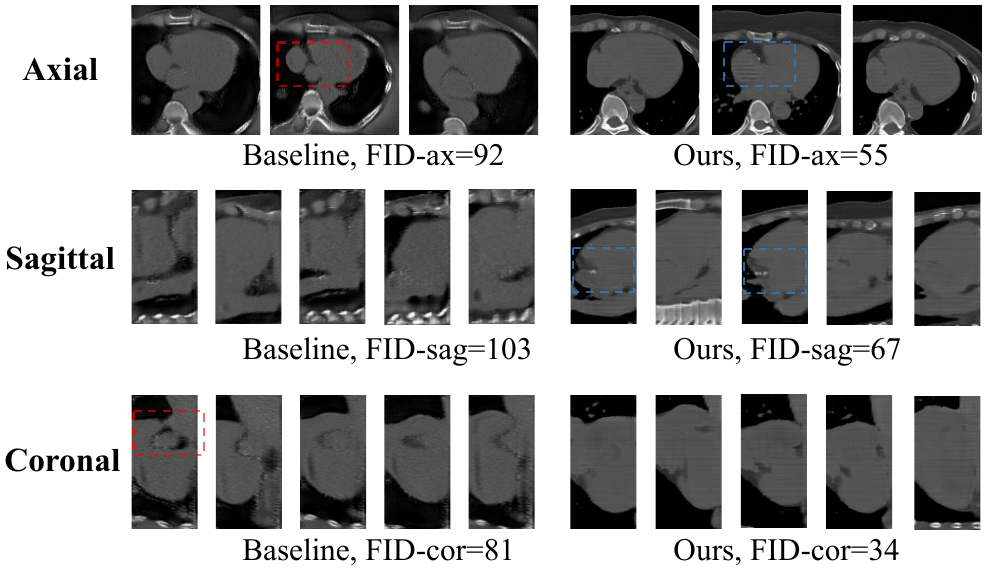}}
\caption{\label{fig:visualization128}Generated images of higher resolution ($128\times 128$) from the Baseline and our Split\&Shuffle method. 
Baseline sometimes shows {anatomically} inconsistent regions (marked in red boxes). 
Our method can generate images with better anatomic structure and possible calcium slices (marked in blue boxes), which will be helpful for downstream tasks.
} 
\end{figure}

\subsubsection{Architecture} 
Our next step was to conduct experiments to comprehensively investigate efficient 3D architectural designs. 
At first, all model variants, including the baseline, are trained from scratch without weight inflation. The FID values and the number of parameters are shown in Table~\ref{tab:architecture}. The models with a ``-D'' suffix apply efficient modules only on the discriminator, thus only leading to a slightly reduced parameter number (e.g., 0.434M vs. 0.600M). ``Group-D'', ``Depthwise-D'', and ``Split\&Shuffle-D'' have a different number of parameters because the architectures are different. 
By contrast, the proposed ``Split\&Shuffle'' design reduces more than half of the parameters compared to the baseline (0.291M vs 0.600M).

With regards to generation performance, the proposed ``Split\&Shuffle'' architecture {achieved} the better FID-avg with the least number of parameters, proving its efficiency and effectiveness. 
All the ``-D'' models produce similar results to the baseline, indicating that modifying just the discriminator has little influence on the generation quality. Note that owing the same number of parameters, ``Split'' performs much worse than the ``Split\&Shuffle'' design. As a result, the Channel Shuffle in Figure~\ref{fig:module} plays a crucial role in ensuring performance.

Then, we {examined} how efficient architectures can be combined with inflation strategies for further performance enhancement. 
Specifically, we {adopted} the best inflation strategy ``Inflate-1'' and {applied} it to all the models in Table~\ref{tab:architecture}. The results are shown in Table~\ref{tab:infArc}.
In this experiment, each individual model has its own pre-trained 2D weights due to the differences in architecture. 
Table~\ref{tab:infArc} demonstrates that all models {achieved} good FID-2D values, once again verifying the rationale for inflating informative 2D weights {to train} 3D GANs. 
Since the number of 2D image slices is sufficient for 2D pre-training, ``Baseline'' with the largest number of parameters {achieved} the best FID-2D. 
However, for the final 3D training, our ``Split$\&$Shuffle'' model achieves the best performance (FID-avg) with the least number of parameters. \emph{\textbf{Compared with the ``Baseline'' model, ``Split$\&$Shuffle'' reduces the FID-avg by 14.7 with only 48.5\% of the parameters.}} 
As {for} the discriminator-only variants with the suffix ``-D'', they {were} much worse than the ``Inflate-1'' baseline. Without channel shuffle operation, ``Split'' {achieved} the worst performance, again showing the indispensable role of channel shuffle in our architecture design. 

\subsubsection{Image Resolution}

\begin{table}[h]
\centering
\caption{\label{tab:highres}Model performance comparison of higher resolution ($128\times 128$) images on COCA dataset. Baseline is the same as Table~\ref{tab:architecture}.}
\setlength{\tabcolsep}{6pt}
\resizebox{0.45\textwidth}{!}{
\begin{tabular}{l|l|lll|l}
\hline
Model                 & \#param & FID-ax & FID-sag & FID-cor & FID-avg \\ \hline
Baseline              & 2.628M & 92  & 103 & 81  & 92.0 \\
Split\&Shuffle        & 1.414M & \textbf{55} & \textbf{67} & \textbf{34}   & \textbf{52.0} \\
\hline 
\end{tabular}
}
\end{table}

To show {that} our method can generate high-resolution images for practical application, we {increased} the resolution of COCA to $128\times 128$. 
We set the base layer to $1\times 4 \times 4$, followed by six upsampling stages. The model capacity {was} also increased by using 64 convolution channels. 

The quantitative results are shown in Table~\ref{tab:highres}. Compared with the Baseline (same as Table~~\ref{tab:architecture}), Split\&Shuffle achieves much better performance with fewer model parameters. 
The visualization of the generated image slices is shown in Figure~\ref{fig:visualization128}. The generated image slices by our method show a more feasible heart anatomy structure and higher image quality. 
In addition, since COCA contains coronary calcium, our method {generated} image slices with possible calcium, which are more realistic. 


\begin{table}[t]
\centering
\caption{\label{tab:sota_ct}Comparison with state-of-the-art methods on COCA (heart) dataset.}
\setlength{\tabcolsep}{6pt}
\resizebox{0.48\textwidth}{!}{
\begin{tabular}{l|l|lll|l}
\hline
Model                 &  \#param & FID-ax & FID-sag & FID-cor & FID-avg \\ \hline
3D-WGAN-GP            & 1.399M  & 191  & 223 &  191  & 201.7 \\
3D-VAE-GAN            & 6.389M & 196  & 281 &  272  & 249.7 \\
3D-$\alpha$-GAN       & 2.811M   & 91   & 88  &  80   & 86.3 \\ 
3D-$\alpha$-WGAN-GP   & 2.811M   & 70   & 86  &  69   & 75.0 \\
3D-StyleGAN2          & 0.600M   & 104  & 92  &  112  & 102.7 \\
Split\&Shuffle (Ours) & \textbf{0.291M} & \textbf{46} & \textbf{66} & \textbf{63} & \textbf{58.3} \\\hline 
\end{tabular}
}
\end{table}

\begin{table}[t]
\centering
\caption{\label{tab:sota_brain}Comparison with state-of-the-art methods on ADNI (brain) dataset.}
\setlength{\tabcolsep}{6pt}
\resizebox{0.48\textwidth}{!}{
\begin{tabular}{l|l|lll|l}
\hline
Model                 &  \#param & FID-ax & FID-sag & FID-cor & FID-avg \\ \hline
3D-WGAN-GP            & 1.817M  & 161 & 161 & 231 & 184.3  \\
3D-VAE-GAN            & 11.001M & 167 & 122 & 224 & 171.0 \\
3D-$\alpha$-GAN       & 3.635M  & 73  & 77  & 99  & 83.0 \\ 
3D-$\alpha$-WGAN-GP   & 3.635M  & 74  & 72  & 87  & 77.7 \\
3D-StyleGAN2          & 0.633M  & 114  & 89 & 99  & 100.7 \\
Split\&Shuffle (Ours) & \textbf{0.325M} & \textbf{65} & \textbf{67} & \textbf{82} & \textbf{71.3} \\\hline 
\end{tabular}
}
\end{table}

\begin{table}[h!]
\centering
\caption{\label{tab:psnr_msssim}PSNR and MS-SSIM evaluations on ADNI (brain) and COCA (heart) datasets. PSNR {was} calculated between real and generated images to evaluate generation quality {(higher values are better)}. MS-SSIM {was} calculated on pairs of generated images to evaluate generation diversity (lower values are better).} 
\setlength{\tabcolsep}{6pt}
\resizebox{0.48\textwidth}{!}{
\begin{tabular}{l|ll|ll}
\hline
                      & \multicolumn{2}{c|}{COCA (heart)} & \multicolumn{2}{c}{ADNI (brain)}  \\\hline
Model                 &  PSNR $\uparrow$ & MS-SSIM $\downarrow$ & PSMR $\uparrow$ & MS-SSIM $\downarrow$ \\ \hline
3D-WGAN-GP            & 16.22  & 0.9988  & 27.11  & 0.9945   \\
3D-VAE-GAN            & 16.38  & 0.8436  & 27.03  & 0.9719  \\
3D-$\alpha$-GAN       & 16.19  & 0.8288  & 27.42  & 0.8750   \\ 
3D-$\alpha$-WGAN-GP   & 15.82  & 0.7916  & 27.71  & 0.8678   \\
3D-StyleGAN2          & 16.04  & 0.8362  & 27.80  & 0.8999   \\
Split\&Shuffle (Ours) & \textbf{16.56} & \textbf{0.7513} & \textbf{28.57} & \textbf{0.8488}  \\\hline 
\end{tabular}
}
\end{table}

\subsection{Comparison with State-of-the-art Methods} 
Finally, we {compared} the performance of our method with the published 3D generative models in Table~\ref{tab:sota_ct} (COCA) and Table~\ref{tab:sota_brain} (ADNI). 
The comparison methods {included} the following 3D generation baselines:
\begin{itemize}
    \item 3D-WGAN-GP~\citep{ref:3DWGAN_GP}, a 3D extension of Wasserstein GAN with Gradient Penalty to alleviate training instability. 
    \item 3D-VAE-GAN~\citep{ref:3DVAEGAN}, consisting of an encoder, a decoder and a discriminator. 
    \item 3D-$\alpha$-GAN~\citep{ref:3Dalpha-GAN}, applying the code discriminator and encoder on top of the conventional GANs to alleviate the collapse and blurriness. 
    \item 3D-$\alpha$-WGAN-GP~\citep{ref:3dvae}, utilizing both the $\alpha$-GAN structure and the WGAN-GP loss. 
    \item{3D-StyleGAN2}~\citep{ref:3dstylegan}, a direct 3D extension of the 2D StyleGAN2 model. 
\end{itemize}

On both the heart and brain datasets, the proposed method outperforms all the state-of-the-art methods by a large margin, demonstrating its effectiveness. 
Among all comparison methods, the first four baselines have a much greater number of parameters than our method but {achieved} inferior performance. 
Although 3D-StyleGAN2 has approximately twice as many parameters as our method, it still {performed} much worse. 
Because the 3D medical {image} generation lacks sufficient 3D training images, most baselines are short of sufficient 3D training images, leading to inferior performance. 
As an alternative, our method uses both the weight inflation and Split\&Shuffle designs to mitigate the reliance on a large number of 3D images for training. 

\begin{figure}[t]
\centerline{\includegraphics[width=0.38\textwidth]{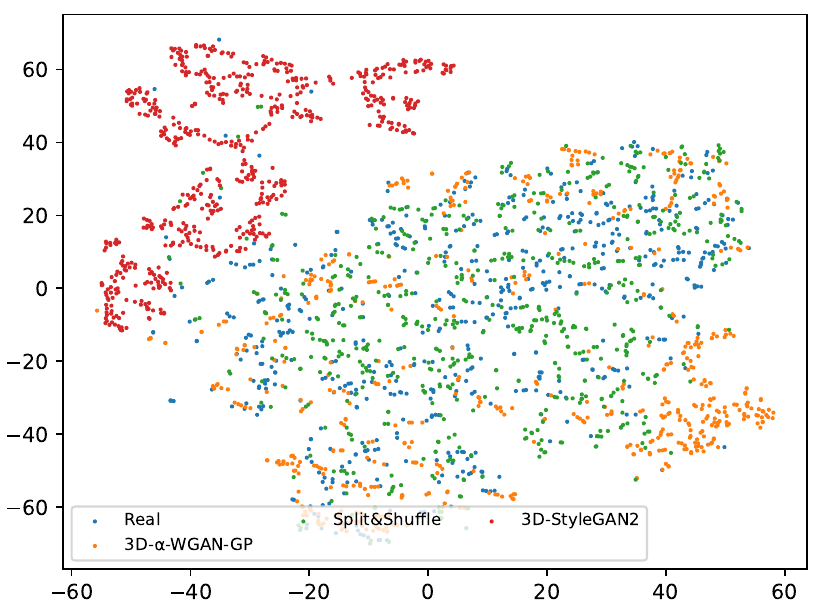}}
\caption{\label{fig:TSNE_CT}t-Distributed Stochastic Neighbour Embedding on COCA dataset.} 
\end{figure}

\begin{figure}[ht]
\centerline{\includegraphics[width=0.38\textwidth]{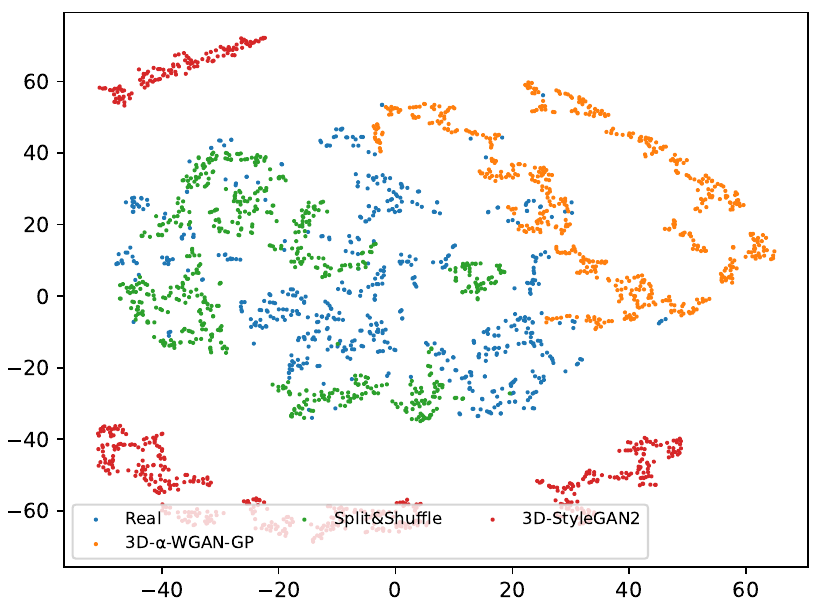}}
\caption{\label{fig:TSNE_MRI}t-Distributed Stochastic Neighbour Embedding on ADNI dataset.} 
\end{figure}

\begin{figure}[t]
\centerline{\includegraphics[width=0.45\textwidth]{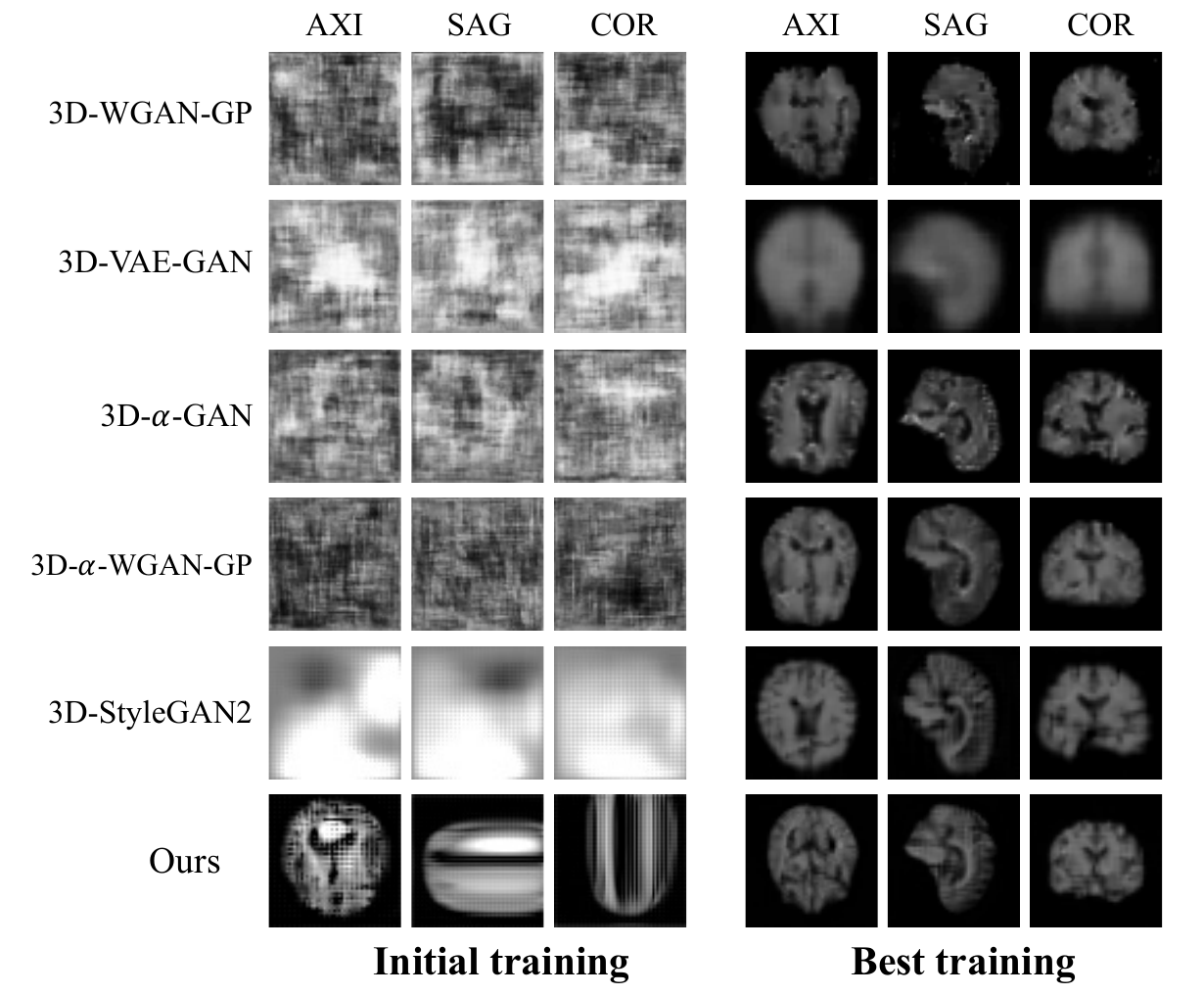}}
\caption{\label{fig:ADNI}Generated images from the initial training and the best training on the brain ADNI dataset. Our method shows good initial brain anatomy, especially for the axial view.} 
\end{figure}

\textbf{Generation quality and generation diversity} 
Except for FID scores, we {considered} two widely-used evaluation metrics: PSNR and MS-SSIM. 
Specifically, PSNR is calculated between the real and generated images to evaluate the generation quality. 
Following \cite{ref:3dvae}, MS-SSIM is calculated on pairs of generated images to evaluate the generation diversity (a smaller value means better diversity). 
The results are shown in Table~\ref{tab:psnr_msssim}. Our method achieved the best PSNR and MS-SSIM performance on both the heart and brain datasets, demonstrating that our method can generate high-quality 3D medical images with better diversity. We also note that the PSNR scores on the heart dataset are smaller than that on the brain dataset. This is due to the large variations among the unaligned heart images. In contrast, the brain images are aligned and exhibit smaller variations.

\textbf{t-Distributed Stochastic Neighbour Embedding (t-SNE)} 
To better understand the distributions of generated and real images, we performed t-SNE on real images, our method, 3D-$\alpha$-WGAN-GP, and 3D-StyleGAN2. 
The visualization of the COCA dataset is shown in Figure~\ref{fig:TSNE_CT}. Although the distributions of both our method and 3D-$\alpha$-WGAN-GP approach the real images, our method {is closer to the} real images. The distribution of 3D-StyleGAN2 is far from {the} real images, consistent with its large FID score. 
As to the visualization of the ADNI dataset in Figure~\ref{fig:TSNE_MRI}, only our method shows {a} similar distribution {to} the real images.

Figure~\ref{fig:ADNI} shows examples of brain slices generated using all comparison methods. 
Our method generates high-quality brain slices on three planes. 
Because of the random weight initialization, the comparison methods generate random images at the start of the training process without any meaningful patterns. 
Due to its training on sufficient 2D axial slices, our model exhibits good anatomy right at the beginning. 
Combining our Split\&Shuffle design with this anatomy prior allows our model to generate better results with fewer parameters. 


\section{Conclusion}

\label{sec:conclusion}
The purpose of {our study was} to address the important problem of generating reliable synthetic 3D medical images. 
The lack of annotated 3D data and inefficient parameter settings hinder the effective training of 3D medical generative models. 
A novel GAN model (i.e., 3D Split\&Shuffle-GAN) is proposed to remedy these problems from two perspectives: training strategy and network architecture. 
For the training strategy, we {used} the weight inflation technique to pre-train a 2D GAN model and inflate the 2D convolution weights as a favorable method for initializing a 3D GAN model. 
For network architecture, we {devised} parameter-efficient Channel Split\&Shuffle modules for the discriminator and generator of the GAN. 
We conducted comprehensive experiments to determine the best weight inflation variant and network architecture design.
The effectiveness of our method is verified on both the heart and brain datasets. 
Further exploration of network weight initialization strategies beyond inflation and the design of new architectures will be {completed} in the future. 

\section{Acknowledgements}


\subsection*{Declaration of Competing Interest}
The authors declare the following financial interests/personal relationships which may be considered as potential competing interests:
Girish Dwivedi reports a relationship with Artrya Pty Ltd. that includes: consulting or advisory and equity or stocks.


\subsection*{Funding}
This work was supported by MRFF Frontier Health and Medical Research - RFRHPI000147. 
The computation in this work was partly supported by DUG facilities via Harry Perkins.

\bibliographystyle{model2-names.bst}\biboptions{authoryear}
\bibliography{main_CMPB_v2}



\end{document}